\documentstyle[aps,epsf]{revtex}

\newcommand{\infig}[2]{\begin{center}\mbox{\epsfxsize #2
\epsfbox{#1}}\end{center}}

\begin{document}

\title{Experimental demonstration of quantum-correlations over more than 10 kilometers}
\draft

\twocolumn[\hsize\textwidth\columnwidth\hsize\csname 
@twocolumnfalse\endcsname

\author{W. Tittel, J. Brendel, B. Gisin, T. Herzog, H. Zbinden, and N. Gisin}

\address{University of Geneva, Group of Applied Physics, 20, Rue de l'Ecole de M\'edecine, CH-1211 Geneva 4, 
Switzerland\\
e-mail: wolfgang.tittel@physics.unige.ch}
\date{\today} 

\maketitle

\begin{abstract}
{Energy and time entangled photons at 
a wavelength of 1310 nm are produced by parametric downconversion in a $\mbox{KNbO}_3$ crystal
and are sent into all-fiber interferometers using a telecommunications fiber network. The two 
interferometers of this Franson-type test of the Bell-inequality are located 10.9~km apart 
from one another. Two-photon fringe 
visibilities of up to 81.6\% are obtained. These strong nonlocal correlations 
support the nonlocal predictions of quantum mechanics and provide evidence that entanglement between 
photons can be maintained over long distances.
} 
\end{abstract}

\vskip1pc
]

Since the foundation of quantum mechanics, the often counterintuitive predictions of 
the theory have continued to puzzle physicists. The most peculiar aspect is entanglement,
whereby, for example, a two-particle system is in a pure state, but each particle 
separately is in a complete mixture. Such entanglement leads to predictions 
for correlation measurements that violate the well-known Bell inequalities 
\cite {bell} and thus cannot 
be explained by local-hidden-variables theories. Polarization \cite{exp,switch}, 
position and momentum 
\cite{rarity} as well as energy and time \cite{brendel92,kwiat93,tapster} 
entangled photons have been used to 
show violations of these inequalities and therefore confirm the strong correlations predicted by 
quantum mechanics. Nevertheless, the nonlocal collapse of the wave function by a "spooky 
action at a distance'' \cite{Einstein} is still a subject for investigation 
\cite{weihs}. 

In this paper we present an experiment on entangled pairs of photons physically
separated by more than 10 km with the source roughly at the center. 
This corresponds to an increased distance by about three orders of 
magnitude over all previous experiments, including the one by Tapster et al.\cite{tapster}.
Our experiment took advantage of already installed optical fibers used for modern optical
telecommunications. The attenuation in such fibers has been drastically improved over the past 
two decades, from several decibel per meter down to 0.35 dB per kilometer. Chromatic dispersion,
another potentially limiting phenomenon for our experiment, has also been optimized for
telecommunications purposes. Standard fibers have zero chromatic dispersion at 1310 nm. 
However, the use of telecommunications fibers has a drawback:
photon counters at telecommunications wavelengths are relatively inefficient and
noisy. Nevertheless, the experimental results clearly indicate that the strong 
quantum correlations can be maintained over such large distances. 

The experimental fact that quantum correlations are maintained over long distances
is interesting for fundamental physics as well as for potential applications. Among the
latter, the most promising one is quantum cryptography \cite{Qcrypto}. 
More futuristic applications include dense coding \cite{densecoding}, 
quantum teleportation \cite{teleportation,expteleportation}, and, more generally, quantum networks and other 
quantum information processing. All these applications require that the quantum systems
(qubits) are protected from environment-induced decoherence. In this respect, the 
present results are very encouraging, at least for photon pairs in optical fibers.
To conclude this introduction, let us mention two of the fundamental issues. First there is the 
local-hidden-variable program. It seems clear that if there are no hidden variables after 10 m,
there are also none after 10 km. Since our experiments improve the physical distance at 
the cost of higher losses, less efficient and more noisy detectors, it is not a better test 
of local-hidden-variables. However, it opens the route for an experiment in which the settings of 
the analyzers can be changed during the flight of the particles and therefore could close the
locality loophole \cite{Locloophole}. Such an experiment has already been made  
\cite{switch}; however, the switching was not really random  \cite{zeilinger86}.
A second fundamental issue is environment-induced decoherence \cite{ZurekPhysTod} and 
"spontaneous collapses" \cite{furry}.
The main effect of decoherence is to spread quantum correlations between the two photons of our
experiment into a formidable entanglement between the photons and the environment. Since the
correlations are not really broken but merely hidden, the coherence can sometimes be rebuilt as in
spin echoes \cite{spinechoe} or, in an example closer to our experiment, using Faraday mirrors to 
reflect light from the end of an optical fiber \cite{plugandplay}. In contrast, the main
effect of spontaneous collapses would be an irreversible  destruction of the coherence
\cite{CSL}. Whether this requires a 
modification of the standard quantum dynamics is controversial \cite{RespZurekPhysTod}. Hopefully, 
developments following ours will lead to experimental tests \cite{Percival}.

In our Franson type experiment \cite{franson89} 
each one of the two entangled photons is directed into an
unbalanced interferometer. The physical distance between these two devices is about 10 km
(the shortest distance between the source and an analyzer is 4.5 km). 
Since the path-length difference of the interferometers, exactly the same for both of them, 
is much greater than the coherence length of the single photons, no second-order interference 
(i.e. no single photon interference)
can be observed. However, due to the entanglement, the possibility of the two photons 
choosing particular outputs can be affected by changing the phase difference in either 
interferometer  ($\delta_1$ or $\delta_2$ respectively). 
This effect is described as fourth-order interference (i.e. two-photon interferences) between the 
probability amplitudes corresponding to two possibilities:
The correlated photons choose both short arms or both long arms through the 
interferometers. Due to the two 
remaining possibilities, the photons choose different arms, the visibility is limited to 
50\%. However, using a fast 
coincidence technique \cite{brendel91}, the latter events can be excluded from registration, 
thus increasing the maximum visibility up to 100\%, and leading to the normalized coincidence 
probability \cite{brendel92}
                        
\begin{equation}  
P_{i,j}:=\frac{1}{4}\Big(1+ijV exp\Big[-\Big(\frac{\lambda(\delta_1-\delta_2)}{2\pi 
L_c}\Big)^2\Big]\cos(\delta_1+\delta_2)\Big),
\label{one}
\end{equation}
\noindent
where $i,j =\pm1$ and $P_{+-}$ is e.g., the coincidence probability between the 
detector labeled + at interferometer 1 and the 
one labeled - at interferometer 2 (see Fig. 1). Experimental deviations from the  maximum 
visibility of 1 are described by the 
visibility factor V. Unequal path-length differences in both interferometers 
$\delta_1-\delta_2\neq0$ increase the possibility to 
differentiate between the long-long and short-short paths (by looking at the differences in 
photon arrival time), and hence 
reduce the visibility. Therefore, the interferogram will always show an envelope representing 
the single-photon coherence length $L_c$.

The schematic setup of the experiment is given in Fig.~2. Light from a single longitudinal mode 
laser diode (RLT6515G; 8~mW at 655.7~nm) passes through a dispersion prism P to separate out the 
residual infrared fluorescence light and is focused into a 
$\mbox{KNbO}_3$ crystal. The crystal is 
oriented to ensure degenerate collinear type-I phase matching for signal and 
idler photons at  1310~nm.
Due to these phase-matching conditions, the single photons exhibit 
rather large bandwidths of 
about 90~nm full width at half maximum (FWHM). Behind the crystal, the pump is separated out by a filter F (RG~1000) while 
the passing down-converted 
photons are focused (lens L) into one input port of a standard 3-dB fiber coupler. 
Therefore half of the pairs are split and exit the source by different 
output fibers. In contrast to previous experiments, the
whole source including stabilization of laser current and temperature is of  much smaller 
dimensions (a box of about $40\times45\times15$ $cm^3$) and hence 
can easily be used outside the laboratory.  
In our experiment, the two-photon source is placed at a telecommunications station near Geneva downtown. 
One of the correlated 
photons travels through 8.1 km of installed standard telecom fiber to an analyzer that is 
located in a second station in 
Bellevue, a little village 4.5 km north of Geneva. Losses are about 5.6 dB. Using 
another installed fiber of 9.3 km, we send the other 
photon to a second analyzer, situated in a third station in Bernex, another little village
about 7.3 km southwest of Geneva and 10.9 km from Bellevue. Absorption in this fiber is 
around 4.9 dB, leading to overall 
losses in coincidences of about a factor of 10. Since the bandwidth of the single photons 
is rather large, we have to 
consider chromatic dispersion effects in the connecting  fibers. From 
measured differential group delays we calculate the 
introduced time jitter at a pump wavelength of 655.7~nm to be about 400~ps FWHM.

\begin{figure}[b]
\infig{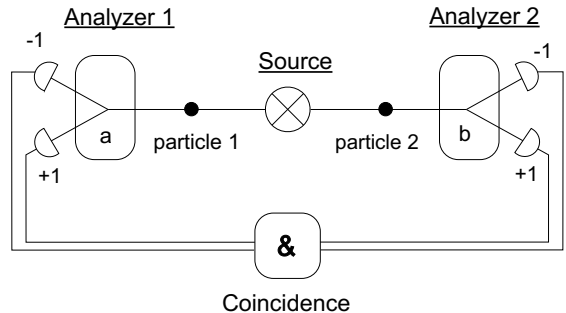}{0.85\columnwidth}                             
\caption{Principle of the setup to test nonlocal correlations. The entangled particles, created by 
         a two-particle source, are separated and each one is sent to an analyzer, matched to 
         the correlated feature (in our case energy-time entanglement and interferometers). 
         Each measuring device, characterized by a setting parameter (a and b respectively), 
         performs a Bernoulli experiment with binary valued output. The results are compared, 
         thus giving evidence to the nonlocal correlations.} 
\label{fig1} 
\end{figure}
\noindent

The two analyzers consist of all-fiber optical Michelson interferometers made of standard 3-dB 
fiber couplers with chemically 
deposited end mirrors. The optical path-length differences (20 cm of optical fiber
or a 1-ns time difference) are equal in
both interferometers. To ensure maximum visibility, birefringence in the fibers 
forming the interferometers has to be avoided. This can 
be attained either by compensating all birefringence effects using Faraday mirrors 
\cite{plugandplay,tittel} or by avoiding birefringence at 
all. In this experiment we aimed to achieve the latter. To do so, we placed the interferometers 
straight and without stress 
into copper tubes. The temperature of the tubes can be varied in order to change and 
control the phase difference of the 
interferometers. To detect the photons, we use passively quenched germanium 
avalanche photodiodes (APD) (NEC NDL5131P1) operated in Geiger mode at 77 K. 
They are biased around 0.4~V above breakdown, leading to a detection-time jitter of 200 ps FWHM, 
an efficiency of about 15\% and dark counts of 100 and 110 kHz each.  
Each single-photon detector triggers 
a 1310-nm laser emitting 1-ns pulses. The (now classical) optical signals are transmitted back 
to Geneva using additional 
fibers and are detected by conventional p-i-n photodiodes. The overall time jitter is about 
450~ps FWHM and permits one to 
differentiate between photons having traveled along different interferometer arms. 
The signals from the p-i-n photodiodes trigger a time to pulse 
height converter (TPHC) (EG\&G Ortec 457), which is placed next to the source in Geneva. 
A window discriminator 
counts coincidences within a 400 psec interval that is matched to the arrival time of 
the two correlated photons having passed 
equivalent paths in the interferometers (either short-short or long-long). Therefore, 
photon-pair detection is limited to the interfering processes only.

\begin{figure}[htb]
\infig{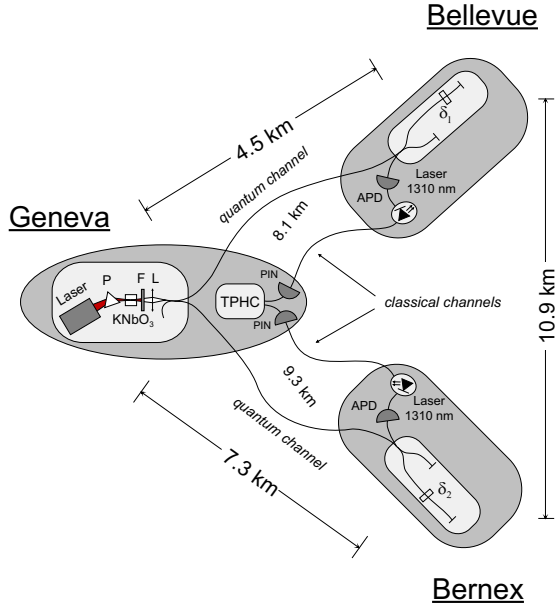}{0.85\columnwidth}
\caption{Experimental setup. See the text for a detailed description.}
\label{fig2}
\end{figure}
\noindent

We monitored the count rates as a function of phase differences $\delta_1,\delta_2$ in both 
interferometers. No phase-dependent variation of the 
single count rates could be observed. They remained constant at about 164 and 167~kHz 
(including 100- and 110-kHz dark counts) for each
detector. Figure~3 shows a plot of the coincidence counts we obtained during intervals of  
20 sec as a function of the 
temperature-caused phase change in the interferometer in Bernex. Applying a Fourier transform 
to these data, we find no more than one frequency exceeding the noise floor, hence confirming 
the hypothesis of the sinusoidal function (Eq.~\ref{one}). Fitting the two-photon interference with 
such a function (Eq.~\ref{one}) 
we get a visibility of V=46\%. This relatively low visibility is due to the high number
of accidental coincidences. The latter are themselves due to simultaneous (i.e. within the 400-psec
coincidence window) dark counts in both detectors or to one
photon of a pair detected simultaneously with a dark count of the other detector, while the
other photon was absorbed by the fiber. 
A quantitative evaluation of these effects is difficult because of the dead time 
($\approx 4 \mu sec$) of the TPHC and of afterpulses.
The total accidental counts was measured by 
introducing an additional delay line, thus placing the coincidence peaks apart from the 
discriminator window. We obtained an average of 150 coincidences per 20 sec \cite{Ames}. 
Subtracting these, the obtained visibility is $(81.6\pm1.1)\%$ \cite{vis}. 
This is the visibility one would expect for noiseless detectors.

\begin{figure}
\infig{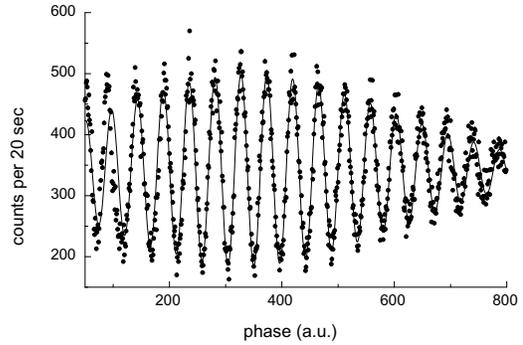}{0.85\columnwidth}
\caption{Net coincidence counts per 20 sec as a function of path-length difference in the Bernex 
         interferometer. A best fit with a sinusoidal function yields a visibility of 
         $(81.6\pm1.1)\%$. From the envelope of the interferogram, we calculate a single-photon 
         coherence length around $10.2\;\mu\mbox{m}$ (equivalent to a bandwidth of about 90~nm FWHM).}
\label{fig3} 
\end{figure}

\noindent
Since we found the same visibility in a "short distance"
Bell experiment in our laboratory, we 
assume that the deviation from the maximum theoretical value of 1 is 
not due to the physical distance between the interferometers. The missing 18\% has to be 
traced back to the imperfect
performance of our devices. Remaining 
birefringence and chromatic dispersion effects in the interferometers, unequal transmission 
probabilities for the different 
arms, and imperfect suppression of the unwanted satellite coincidence peaks are possible 
explanations. However, the main 
reduction can probably be found in small wavelength fluctuations of the pump laser. 

As mentioned in the introduction, this experiment was not primarily designed to test local
hidden variables. However, it is interesting to evaluate by how much the Bell inequality
could be violated in such a long distance Bell experiment. This inequality
provides also a natural
criterion for a "strong quantum correlation": The correlation is strong if the two-photon
interference visibility is larger than the 71\% necessary to infer a violation of Bell's
inequality. It is remarkable that this definition of "strong quantum correlation" corresponds
also precisely to the visibility necessary in quantum cryptography to guarantee the security of the
quantum communication channel: If the perturbations introduced by Eve while trying to eavesdrop
are weak, in the precise sense that the two-photon visibility remains larger than 71\%, 
then one can prove that her optimal information on Alice data is necessarily lower than Bob's
information \cite{EveBell}. Conversely, if the perturbations are strong, i.e. if the two-photon
visibility reduced below 71\%, then Eve's information may be larger than Bob's one.
Looking at the obtained visibility after subtraction of the accidental counts
of $(81.6\pm1.1)\% > 1/\sqrt{2}\approx0.71$, we can infer a violation of the Bell inequality 
by ten standard deviations \cite{Pij}. 

In conclusion, this experiment gives evidence, that the spooky action 
between entangled photons does not 
break down when separating the particles by a physical distance of 10~km. 
However, we acknowledge that neither 
this nor any of the previous experiments can close the detection loophole \cite{detectloophole}. 
Hence in 
order to deduce nonlocality from existing data, additional assumptions are needed. In
particular, and most importantly in our opinion, one has to assume 
that the detected pairs of particles form a fair sample of the set of all emitted pairs. 
Further investigations with interferometers and a 
pump laser of better performance should be made, also varying the fiber lengths. 
In a 
next experiment we plan to test Bell's inequalities with truly random settings of the analyzers in 
order to rule out every subluminal communication between the different experimental parts thus
obeying Einstein locality. In addition to tests of fundamental physics, the possibility of 
creating entangled photons at telecommunications wavelength and the fact that the different experimental 
parts (especially the source) are of small dimensions and easy to handle opens the field for 
future application of photon pairs. More generally, it nicely illustrates a general aim of
the field of quantum information processing: It turns quantum conundrums into potentially
useful processes.

This work was supported by the Swiss Priority Program "Optique" and by the TMR network "The 
physics of Quantum Information", Contract No. ERBFM-RXCT960087. We would like to thank B. Huttner for
helpful discussions, J. D. Gautier for technical support
and the Swiss Telecom for placing the stations and the optical fibers at our disposal.



\begin{references}
\bibitem{Ames} The accidental coincidence count rate was measured over a much longer time than
           the reported value for 20 second. Hence the accuracy of this value is better than
           1\%.
\bibitem{vis}To determine the visibility corresponding to the measured data, we apply a 
             least squares fit ($\chi^2$-test), assuming the sinusoidal function (\ref{one}) with V as 
             the free parameter. The reliability of the best fit can be found by looking at the 
             tolerated variation of the fitted parameters in order to keep the $\chi^2$ value 
             smaller than a certain limit. This threshold, depending on the number of 
             measured data and the desired confidence level (as common in statistics, we 
             choose 0.95) can be found in tables. Thus, the probability that our data are 
             corresponding to a sinusoidal function with a visibility of $(81.6\pm1.1)\%$ is 0.95.
\bibitem{Pij} In order to infer this violation of the Clauser-Horne-Shimony-Holt-type of the Bell 
           inequality, we make the usual
           assumption about the "detection loophole", i.e. we identify probabilities with
           counting frequencies. Moreover, we use a natural symmetry argument to assume
           $P_{+-}(a,b) = 1/2 - P_{++}(a,b)$ and similarly for $P_{--}(a,b)$ and $P_{-+}(a,b)$.
\end{references}
\end{document}